\documentclass[preprint2]{aastex}
\usepackage{spr-astr-addons}
\usepackage{graphicx}
\usepackage{psfig}
\newcommand{\be}{\begin{equation}}
\newcommand{\ee}{\end{equation}}

\newcommand{\ledd}{$L_{\rm Edd}$~}

\newcommand{\gtsima}{$\; \buildrel > \over \sim \;$}
\newcommand{\ltsima}{$\; \buildrel < \over \sim \;$}
\newcommand{\prosima}{$\; \buildrel \propto \over \sim \;$}
\newcommand{\gsim}{\lower.5ex\hbox{\gtsima}}
\newcommand{\lsim}{\lower.5ex\hbox{\ltsima}}
\newcommand{\simgt}{\lower.5ex\hbox{\gtsima}}
\newcommand{\simlt}{\lower.5ex\hbox{\ltsima}}
\newcommand{\simpr}{\lower.5ex\hbox{\prosima}}

\newcommand{\etal}{{et al.~}}
\newcommand{\cxo}{\textit{Chandra~}}
\newcommand{\spi}{\textit{Spitzer~}}

\begin{document}
\title{
Galactic X-ray binary jets}

\shorttitle{Galactic X-ray binary jets}       
\shortauthors{Gallo}

\author{Elena Gallo}
\affil{Chandra Fellow\\University of California, Santa Barbara, CA 93106, USA}


\begin{abstract} 
 
\noindent
With their relatively fast variability time-scales, Galactic X-ray binaries provide an
excellent laboratory to explore the physics of accretion and related
phenomena, most notably outflows, over different regimes. 
After comparing the phenomenology of jets in black hole systems 
to that of neutron stars, here I discuss 
the role of the jet at very low Eddington ratios, and present preliminary
results obtained by fitting the broadband spectral energy distribution of a 
quiescent black hole binary with a `maximally jet-dominated' model. 
\end{abstract}





\section{Black holes}
\label{sec:bhs}
A key observational aspect of X-ray binary jets is their synchrotron
radio emission.
In black hole systems, radiatively inefficient, hard X-ray states (McClintock
\& Remillard 2006; Homan \& Belloni 2005) are
associated with flat/slightly inverted radio-to-mm spectra and persistent
radio flux levels (Fender 2001). In analogy with compact extragalactic radio
sources (Blandford \& K\"onigl 1979), the flat spectra are thought to be due to the
superposition of a number of peaked synchrotron spectra generated along a
conical outflow, or jet, with the emitting plasma becoming progressively
more transparent at lower frequencies as it travels away from the jet base.  The jet
interpretation has been confirmed by high resolution radio maps of two hard
state black hole binaries (BHBs): Cygnus X-1 (Stirling \etal 2001) and GRS~1915+105 (Dhawan
\etal 2000) are both
resolved into elongated radio sources on milliarcsec scales -- that is tens of
A.U. -- implying collimation angles smaller than a few degrees. Even though no
collimated radio jet has been resolved in any BHB emitting X-rays below a few
per cent of the Eddington limit, it is widely accepted, by analogy with the
two above-mentioned systems, that the flat radio spectra associated with
unresolved radio counterparts of X-ray binaries are originated in conical
outflows. 

Radiatively efficient, thermal dominant (high/soft) X-ray states, on the
contrary, are associated with no detectable core radio emission (Fender
\etal 1999); as the radio fluxes drop by a factor up to 50 with respect to the
hard state (e.g. Corbel \& Fender 2002, Corbel \etal 2004), this is generally
interpreted as the physical suppression of the jet taking place over this
regime.

Transient ejections of optically thin radio plasmons moving away from the
binary core in opposite directions are often observed as a result of bright
radio flares associated with hard-to-thermal X-ray state transitions. As
proven by the case of e.g. GRS~1915+105 (Mirabel \& Rodr\'\i guez 1998), the same
source can produce either kind of jets, steady/partially self-absorbed,
and transient/optically thin, depending on the accretion regime.

\subsection{A unified model for black hole X-ray binaries}
\label{sec:uni}

Fender, Belloni \& Gallo (2004) have addressed the issue of whether the steady
and transient jets of BHBs have a different origin or are somewhat different
manifestations of the same phenomenon, showing that: i) the power content of
the steady and transient jets are consistent with a monotonically increasing
function of $L_{X}$; ii) the measured bulk Lorentz factors of the transient
jets seem to be higher than those inferred for the steady jets. Based upon
these arguments, the first unified model for the jet/accretion coupling in
BHBs has been proposed. The key idea is that, as the thin disk inner boundary
moves closer to the hole (hard-to-thermal state transition), the escape
velocity from the inner regions increases.  As a consequence, the steady jet
bulk Lorentz factor rises sharply, causing the propagation of an {\em internal
shock} through the slower-moving outflow in front of it.  One typical cycle
(lasting days to years, depending on the source) can be described as follows:
the system is in the hard X-ray state, producing a steady jet. At some point
$L_{\rm X}/L_{\rm Edd}$ increases above a few per cent while the X-ray
spectrum softens, i.e. the accretion disk inner boundary is moving closer and
closer to the BH, resulting in an increase of the escape velocity.  A sudden
ejection of hot electrons (perhaps coupled with baryons) with high bulk
Lorentz factor causes the propagation of a shock throughout the pre-existing outflow,
and finally disrupts it. Eventually, the result of this shock is what we
observe as a post-outburst, optically thin radio plasmon.  No emission from
the steady jet is detected until the inner disk recedes once more, in which
case a new cycle begins.

However, there are at least a couple of recent results that might
challenge some of the premises the proposed unified scheme is based
on.  The first one is the notion that, for the internal shock scenario
to be at work and give rise to the bright radio flare at the state
transition, whatever is ejected must have a higher velocity with
respect to the pre-existing hard state steady jet. From an
observational point of view, this was supported, on one side, by the
lower limits on the transient jets' Lorentz factors, typically higher
than 2 (Fender 2003), and, on the other hand, by the relative small
scatter about the radio/X-ray correlation in hard state BHBs~(Gallo,
Fender \& Pooley 2003). The latter has been challenged on theoretical
grounds (Heinz \& Merloni 2004), while a recent work (Miller-Jones
\etal 2006) has demonstrated that, from an observational point of
view, the average Lorentz factors do not differ substantially between
hard and transient jets (albeit the estimated Lorentz factors rely on
the assumption of no lateral confinement).

In addition, recent high statistics X-ray observations of hard state BHBs
undergoing outburst (Miller \etal 2006; Rykoff \etal 2007) suggest
that a cool, thin accretion disk extends already near to the innermost stable
circular orbit (ISCO) already during the bright phases of the hard state, that
is prior to the top horizontal brunch in the top panel of Figure 7 in Fender,
\etal (2004).
This would challenge the hypothesis of a sudden deepening of the inner disk
potential well as the cause of a high Lorentz factor ejection.  Possibly,
whether the inner disk radius moves close to the hole prior or during the
softening of the X-ray spectrum does not play such a crucial role in terms of
jet properties; if so, then the attention should be diverted to a different
component, such as the presence/absence, or the size (Homan \etal 2001) of a
Comptonizing corona (which could in fact coincide with the very jet
base; Markoff, Nowak \& Wilms 2005).\\

Finally,
much work needs to be done
in order to test the consistency of the internal shock scenario as a viable
mechanism to account for the observed changes in the radio properties, given
the various observational and theoretical constraints (such as
emissivities, radio/infrared delays, cooling times, mass outflow rates, etc.).

One of the most interesting aspects of this proposed scheme -- assuming that
is correct in its general principles -- is obviously its possible application
to super-massive BHs in Active Galactic Nuclei (AGN), and the possibility to
mirror different X-ray binary states into different classes of AGN: radio loud
vs. radio quiet, LLAGN, FRI, FRII etc.. The interested reader is referred
to~K\"ording, Jester \& Fender (2006).

\section{Neutron stars}
\label{sec:ns}
Low magnetic field, `atoll-type' neutron star X-ray binaries (see van der Klis
2006 for a classification) share many X-ray spectral and timing properties
with BHBs and show two distinct X-ray states, which can be directly compared
to the hard and thermal state of BHBs: the reader is referred to Migliari \&
Fender (2006) for a comprehensive work of correlated radio and X-ray
properties of neutron stars (NSs). The main differences/similarities can be
summarized as follows: as for the BHBs, below a few per cent of their
Eddington luminosity, NSs power steady, self-absorbed jets. Transient
optically thin plasmons are ejected at higher X-ray luminosities, in response
to rapid X-ray state changes. However, {\em NSs are less `radio loud' than
BHBs.} At a given $L_{\rm X} / L_{\rm Edd}$, the difference in radio
luminosity is typically a factor $\sim$30, which {can not be accounted for by
the mass ratio alone}.  In contrast to BHs, atolls have been detected in the
radio band while in the soft X-ray state, suggesting that the jet quenching in
disk-dominated states may not be so extreme.  Like in the BHBs, the radio
luminosity of hard state NSs scales non-linearly with the X-ray luminosity
(see Section~\ref{sec:corr}), even though the slope of radio-to-X-ray
correlation --albeit the low statistics-- appears to be steeper for the NSs
($L_{\rm radio}\propto L_{\rm X}^b$, with $b\simeq 0.7$ for the BHs, and $b\simeq
1.4$ for the NSs; see K\"ording, Fender \& Migliari 2006 for an
interpretation).

High magnetic field, `Z-type' NSs, being persistently close to the Eddington
accretion rate, behave similar to the rapidly varying BHB GRS 1915+105.  The
role played by the NS dipole magnetic field in producing/inhibiting the jet
production mechanism is far from being clear, although it is generally assumed
that the higher the magnetic field, the lower the jet power.  Further radio
observations of X-ray pulsars and ms accreting X-ray pulsars are definitely
needed in order to place stronger observational constraints.\\

Overall, these results indicate that, at the zeroth order, the process of
continuous jet formation 
works similarly in accreting BHs and (at least certain) NSs. Even though
quantitative differences remain to be explained, such as the relative dimness
of neutron star jets with respect to BH jets, this indicates that the presence
of an accretion disk coupled to an intense gravitational field would be the
necessary and sufficient ingredients for steady relativistic jets to be 
formed, regardless of the presence/absence of an event horizon. 
In addition, it is worth reminding that the most relativistic Galactic jet
source observed so far is a NS~(Fender \etal 2004), contrary to what it might
be expected according to the so called `escape-velocity paradigm'.

\section{Luminosity Correlations}
\label{sec:corr}
\begin{figure}[t!]
\begin{center}
\includegraphics[width=0.40\textwidth]{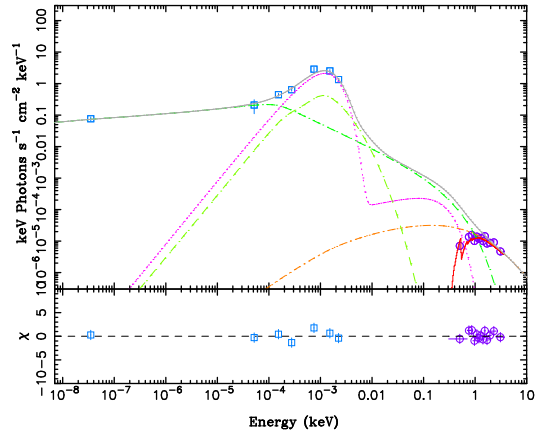}
\end{center}
\caption{Fit to broadband SED of the BHB A0620-00 with the `maximally
jet-dominated model'.  The symbols represent the data, while the solid
red line is the model fit in detector space.  Solid (gray): total
spectrum; Dot-long-dashed (light green): pre-acceleration inner jet
synchrotron emission; Dot-long-dashed (darker green):
post-acceleration outer jet synchrotron; Dot-dash-dash (orange):
Compton emission from the inner jet; Dot-short-dash (magenta): thermal
multicolor-blackbody disk model plus single blackbody representing the
star. From Gallo \etal (submitted to ApJ).}
\label{fig:sed}       
\end{figure}
%
%

{\it Radio/X-ray}\smallskip\\
\noindent
In a first attempt to quantify the relative importance of jet vs. disk
emission in BHBs, Gallo \etal (2003) collected quasi-simultaneous radio and
X-ray observations of ten hard state sources. This study established the
presence of a tight correlation between the X-ray and the radio luminosity, of
the form $L_{\rm R}\propto L_{\rm X}^{0.7\pm 0.1}$, first quantified for
GX~339--4~(Corbel \etal 2003). The correlation extends over more than 3 orders
of magnitude in $L_{\rm X}$ and breaks down around 2 per cent of the Eddington
X-ray luminosity, above which the sources enter the thermal dominant state,
and the core radio emission drops below detectable levels. Given its
non-linearity, the ratio radio-to-X-ray luminosity increases towards
quiescence (i.e. below a few $10^{-6} L_{\rm Edd}$). Thanks to the quite large
degree of uncertainty about the overall structure of the accretion flow in
this regime, this has led to the hypothesis that the total power output of
quiescent BHBs could be dominated by a radiatively inefficient outflow, rather
than by the local dissipation of gravitational energy in the accretion flow
(Fender, Gallo \& Jonker 2003; K\"ording, Fender \& Migliari 2006).
Simultaneous radio/X-ray observation of the nearest quiescent BHB A0620--00
have confirmed that the non-linear correlation holds down to Eddington ratios
as low as $10^{-8}$ (Gallo \etal 2006).  Nevertheless, many outliers have been
recently been found at higher luminosities (see Gallo 2007, Xue
\& Cui 2007, and references therein), casting  
doubts on the universality of this scaling, and the 
possibility of relying on the best-fitting relation for
estimating other quantities, such as distance or BH mass.\smallskip\\
\noindent
{\it Optical-Infrared/X-ray}\smallskip\\
\noindent
The infrared (IR) spectra of BHBs with a low mass donor star are
likely shaped by a number of competing emission mechanisms, most
notably: reprocessing of accretion-powered X-ray and ultraviolet
photons, either by the donor star surface or by the outer accretion
disk, direct thermal emission from the outer disk, and non-thermal
synchrotron emission from a relativistic outflow.
Russell \etal (2006) have collected all the available quasi-simultaneous
optical and near-IR data of a large sample of Galactic X-ray binaries over
different X-ray states. The optical/near-IR (OIR) luminosity of hard/quiescent
BHBs is found to correlate with the X-ray luminosity to the power $\sim$0.6,
consistent with the known radio/X-ray correlation slope down to
$10^{-8}$\ledd~(Gallo \etal 2006).  Combined with the fact that the near-IR
emission is largely suppressed in the thermal-dominant state, this leads to
the conclusion that, for the BHBs, the break to the optically thin portion
would take place in the mid-IR (2-40 $\mu$m).
A similar correlation is found in low-mass neutron stars (NSs) in the
hard state.
By comparing the observed relations with those expected from models of a
number of emission processes, Russell \etal (2006) are able to constrain the
mean OIR contributions to the spectral energy distribution (SED) for the
different classes of X-ray binaries. They conclude that the jets are
contributing 90 per cent of the near-IR emission at high luminosities in the
hard state of BHBs. The optical emission could have a substantial (up to 75
per cent) jet contribution; however, the optical SEDs show a thermal spectrum
indicating X-ray reprocessing in the disk dominates in this regime.
In contrast, X-ray reprocessing dominates the OIR in
hard state NSs, with possible contributions from the jets 
and the viscously heated disk only at high luminosities.

\section{The role of the jet in quiescence}
\label{sec:qjet}

There is evidence from large scale-structures that the jets' mechanical power
is comparable to the bolometric X-ray luminosity in some hard state BHB sources
(e.g. Russell \etal 2007). However,
even for the highest quality SED, disentangling
the relative contributions of inflow vs. outflow to the radiation spectrum and
global accretion energy budget can be quite challenging, as illustrated by the emblematic
case of 
\object{XTE J1118+480} in McClintock \etal (2003) and Markoff \etal (2003).
Estimates of the total jet power based on its radiation spectrum depend
crucially on the assumed frequency at which the flat, partially self-absorbed
spectrum turns and becomes optically thin, as the jet `radiative efficiency'
depends ultimately on the location of the high-energy cutoff induced by the
higher synchrotron cooling rate of the most energetic particles. Once again,
this quantity has proved hard to measure.
Indeed, from a theoretical point of view, the `break frequency',
here defined as the frequency at which the partially self-absorbed jet
becomes optically thin, is inversely proportional to the BH mass: as
jet spectral breaks are often observed in the GHz/sub-mm regime in
active nuclei, they are expected to occur in the IR-optical band for
$10^{5-7}$ times lighter objects.
We know however from observations of \object{GX~339--4}, the only BHB
where the optically thin jet spectrum has been perhaps observed (Corbel \&
Fender 2002, Homan \etal 2005), that the exact break frequency can
vary with the overall luminosity, possibly reflecting changes in the
magnetic field energy density, particle density and mass loading at
the jet base. Determining the location of
the jet break as a function of the bolometric luminosity is important
to assess the synchrotron contribution to the hard X-ray band.  
As an example, that the optically thin jet IR-emission in GX~339--4
connects smoothly with the hard X-ray power law has led to challenge
the `standard' Comptonization scenario for the hard X-ray state~(Markoff \etal
2001), 
whereas recent \spi observations of
the ultra-compact neutron star X-ray binary 4U~0614+091~(Migliari \etal 2006) 
revealed that the break frequency must take place in the
far-IR in this system, effectively ruling out a synchrotron origin for
the X-ray power law.

\subsection{Case study: A0620--00}
\label{a0620}

The role of the jet --if any-- is especially interesting at very low
luminosities, in the so called `quiescent' regime, i.e. below a few 
$10^{-6} L_{\rm Edd}$.
Steady jets appear to survive down to quiescent X-ray
luminosities (Gallo \etal 2006), even though sensitivity limitations
on current radio telescopes make it extremely difficult to reach the
signal-to-noise ratios required to assess their presence for low
luminosity systems further than 2 kpc or so.
In spite of the large degree of uncertainty on the overall geometry of
the accretion flow in this regime, there is general agreement
that the X-ray emission in quiescent BHBs comes from
high-energy electrons near the BH, the disagreement comes about in:
{\it i)} attributing the emission to outflowing vs. inflowing
electrons; {\it ii)} modeling the electron distribution as thermal
vs. non-thermal (or hybrid; McClintock \etal 2003). 
The SEDs of quiescent BHBs, as well as low-luminosity AGN 
are often examined in the context of the advection-dominated accretion
flow solution (Narayan \& Yi 1994), whereby the low X-ray luminosities
would be due to a highly reduced radiative efficiency, and most of the
liberated accretion power disappears into the horizon. Alternatively, 
building on the work by~Falcke \& Biermann (1995) on AGN jets, a
jet model has been proposed for hard state BHBs (see Markoff \etal 2005 for the
latest version).
The model is based upon four assumptions: 1) the total power in the
jets scales with the total accretion power at the innermost part of
the accretion disk, $\dot{m}c^2$, 2) the jets are freely expanding and
only weakly accelerated via their own internal pressure gradients
only, 3) the jets contain cold protons which carry most of the kinetic
energy while leptons dominate the radiation and 4) some fraction of
the initially quasi-thermal particles are accelerated into power-law
tails.

Figure~\ref{fig:sed} shows a fit to the SED of A0620--00, the lowest Eddington-ratio BHB
with a known radio counterpart ($L_{\rm
X}/L_{\rm Edd}\simeq 10^{-8}$) with such a `maximally jet-dominated' model
(Gallo \etal 2007, submitted to ApJ). 
This is the first time that such a complex model is
applied in the context of quiescent BHBs, and with the strong
constraints on the jet break frequency cut-off provided by the \spi
data in the mid-IR regime.  
The model is most sensitive to the fitted parameter $N_{\rm j}$, which
acts as a normalization, though it is not strictly equivalent to the
total power in the jets (see discussion in
MNW05). It dictates the power initially divided
between the particles and magnetic field at the base of the jet, and
is expressed in terms of a fraction of 
$L_{\rm Edd}$.
Once $N_{\rm j}$ is specified and conservation is assumed, the macroscopic
physical parameters along the jet are determined assuming that the
jet power is roughly shared between the internal and external
pressures. 
The radiating particles enter the base of the jet where
the bulk velocities are lowest, with a quasi-thermal
distribution. Starting at location $z_{\rm acc}$ in the jets, a free
parameter, a fraction 85$\%$ of the particles are accelerated into a powerlaw
with index $p$, also a fitted parameter.  
The maximum energy of the accelerated leptons is calculated by setting
the acceleration rate to the local cooling rates from synchrotron and
inverse Compton radiation at $z_{\rm acc}$.  If the acceleration
process is diffusive Fermi acceleration, the acceleration rate depends
on the factor $f=\frac{(u_{\rm acc}/c)^2}{f_{sc}}$, where $u_{\rm acc}$ is
the shock speed relative to the bulk plasma flow, and $f_{\rm sc}$ is
the ratio of the scattering mean free path to the gyro-radius.  Because
neither plasma parameter is known, we fit for their combined
contribution $f$, which thus reflects the {\it efficiency of acceleration}.  
{\it We find $f$ to be around two
orders of magnitude lower for A0620--00 than in higher luminosity
sources. }
This `weak acceleration' scenario is reminiscent of the Galactic
Center super-massive BH Sgr A*. Within this framework, the SED of Sgr
A* does not require a power law of optically thin synchrotron emission
after the break from its flat/inverted radio spectrum (Falcke \& Markoff 2000). Therefore, if
the radiating particles have a power-law distribution, it must be so
steep as to be indistinguishable from a Maxwellian in the optically
thin regime. In this respect, they must be only weakly accelerated.
Here we have shown that something similar, albeit less extreme, is occurring
in the quiescent BHB A0620--00; either scenario implies that acceleration in
the jets is inefficient at $10^{-9}-10^{-8}L_{\rm Edd}$.

\begin{acknowledgements}
This work is supported by NASA through
\cxo Postdoctoral Fellowship grant number PF5-60037, issued by
the \cxo X-Ray Center, which is operated by the Smithsonian Astrophysical
Observatory for NASA under contract NAS8-03060. I wish to thank the organizers
for the interesting conference, as well as for the memorable excursion to Tidbinbilla.
 
\end{acknowledgements}



\end{document}